\documentclass[twocolumn,           
showpacs,            
               preprintnumbers,     
               aps,                 
               prd,          	
               a4paper,             
               nofootinbib,         
               tightenlines,        
               floats               
               ]{revtex4}

\usepackage{graphicx}
\usepackage{amssymb,latexsym}

\newcommand{\ba}{\begin{eqnarray}}
\newcommand{\ea}{\end{eqnarray}}  
\newcommand{\be}{\begin{equation}}
\newcommand{\ee}{\end{equation}}

 \newcommand{\CB}{{\cal B}}
\newcommand{\CC}{{\cal C}}

 \newcommand{\CL}{{\cal L}}
 \newcommand{\CN}{{\cal N}}

 \newcommand{\CZ}{{\cal Z}}
\newcommand{\calP}{{\cal P}}

\newcommand{\bear}{\begin{array}}  \newcommand{\eear}{\end{array}}
\newcommand{\bea}{\begin{eqnarray}}  \newcommand{\eea}{\end{eqnarray}}
\newcommand{\beq}{\begin{equation}}  \newcommand{\eeq}{\end{equation}}
\newcommand{\bef}{\begin{figure}}  \newcommand{\eef}{\end{figure}}
\newcommand{\bec}{\begin{center}}  \newcommand{\eec}{\end{center}}
  
\newcommand{\lmk}{\left(}  \newcommand{\rmk}{\right)}
\newcommand{\lkk}{\left[}  \newcommand{\rkk}{\right]}

\newcommand{\vect}[1]{\mbox{\boldmath${#1}$}}
\newcommand{\vecs}[1]{\mbox{\scriptsize\boldmath${#1}$}}

\newcommand{\la}{\left\langle} \newcommand{\ra}{\right\rangle}

\newcommand{\vex}{\mbox{\boldmath${x}$}}

\newcommand{\vecx}{{\vect x}}

\newcommand{\veck}{{\vect k}}
\newcommand{\vecnt}{\hat{\vect n}}
\newcommand{\vecsx}{{\vecs x}}

\newcommand{\vecsk}{{\vecs k}}

\newcommand{\calrk}{{\cal R}_{\vecs k}(0)}

\begin{document}


\title{ Reconstruction of the primordial fluctuation spectrum 
from the five-year WMAP data \\ 
by the cosmic inversion method with band-power decorrelation analysis }

\author{Ryo Nagata and Jun'ichi Yokoyama}


\affiliation{Research Center for the Early Universe (RESCEU),
Graduate School of Science, The University of Tokyo, Tokyo 113-0033,
Japan }

\date{\today} 


\preprint{RESCEU-60/08}


\begin{abstract}
The primordial curvature fluctuation spectrum is reconstructed 
by the cosmic inversion method using the five-year Wilkinson Microwave Anisotropy Probe data 
of the cosmic microwave background temperature anisotropy. 
We apply the covariance matrix analysis and decompose the reconstructed spectrum 
into statistically independent band-powers. 
The statistically significant deviation from 
a simple power-law spectrum suggested by the analysis of the first-year data is not found 
in the five-year data except possibly at one point near the border of the wavenumber domain where accurate 
reconstruction is possible. 
\end{abstract}

\maketitle

\section{Introduction}
Inflationary cosmology \cite{inf1,inf2,inf3} 
explains the origin of cosmic structure on various scales in an unified way. 
The expansion history during the inflationary stage of the early universe 
is recorded in the spectrum of seed structure which can be revealed by 
modern cosmological observations such as the Wilkinson Microwave Anisotropy Probe (WMAP) observation 
of the cosmic microwave background (CMB) 
\cite{WMAPBASIC,WMAPTEMP,WMAPCOSMO,WMAPINF,WMAPTTTE,loglike,WMAP3TEMP,WMAP5,WMAP5ONLY,WMAP5COSMO}. 
The WMAP mission evaluated every multipole moment of the CMB anisotropy spectrum 
in a wide range of scales 
which is potentially a record of the inflationary expansion history with {\it high time resolution}. 
It is a feasible challenge to probe the highly time-resolved behavior 
of the inflaton field (s) which drives inflation. 

Since the initial release of the WMAP results 
\cite{WMAPBASIC,WMAPTEMP,WMAPCOSMO,WMAPINF,WMAPTTTE}, 
it has been argued that the CMB temperature anisotropy spectrum 
has nontrivial features such as running of the spectral index, 
oscillatory behaviors on intermediate scales, 
and lack of power on large scales \cite{INFRA,JF94,COBE,JY99}, 
which cannot be explained by a  power-law primordial spectrum 
that is a generic prediction of simplest inflation models. 
These features may provide clues to unnoticed physics of inflation. 
Some of these anomalous structures disappeared on the three-year spectrum, 
however several anomalies are still existing \cite{WMAP3TEMP,COSPA07}. 
To explain these features, a number of inflation models have been proposed 
\cite{CPKL03,CCL03,FZ03,KT03,HM04,LMMR04,KYY03,YY03,KS03,YY04,MR04A,MR04C,KTT04,HS04,HS07,CHMSS06}. 

As an observational approach to these possible nontrivial features, 
which can be an alternative to model fitting, 
there have been several attempts to reconstruct the primordial spectrum 
using CMB anisotropy data without any prior assumptions about the shape of the primordial spectrum 
\cite{WSS99,BLWE03,MW03A,MW03B,MW03C,MW03D,SH01,SH04,BLH07,SVJ05,WMAP3COSMO,LSV08,VP08}. 
One such attempt to reconstruct the primordial spectrum is a filtering method where 
the primordial spectrum is characterized by amplitudes on a few number of representative scales. 
While such methods have an advantage for reconstructing 
the global structure of the primordial spectrum, 
they may miss possible fine structures if their scale width 
is narrower than the filtering scale, which has been chosen rather arbitrarily so far. 

On the other hand, there exist other methods which can reconstruct the primordial spectrum 
as a continuous function without any {\it ad hoc} filtering scale to investigate detailed features
such as the cosmic inversion method \cite{MSY02,MSY03,KMSY04,KSY04,KSY05},
the Richardson-Lucy method \cite{SS03,SS06,SS07}, or a nonparametric method \cite{TDS04,THS05}. 
The cosmic inversion method has proved its ability of 
reconstructing the modulations off a power-law spectrum quite well 
by the analysis of mock anisotropy data. 
In the analysis of the first-year WMAP data, we pointed out 
the possibility of nontrivial structures in the primordial spectrum around 
the scales of $2.4\times10^2 {\rm Mpc}$ and $4.2\times10^2 {\rm Mpc}$. 

Theoretically, according to the standard inflation paradigm, 
each wavenumber ($k$-) mode of the power spectrum is mutually independent. 
On the other hand, each $k$-mode of the power spectrum reconstructed 
from the observed CMB anisotropy has a strong correlation with neighboring 
modes because each multipole of CMB anisotropy, $C_\ell$, 
depends on the $k$-modes in the wide range around $k=\ell/d$ where 
$d$ is the distance to the last scattering surface. 
In order to extract real features, therefore, it is important to decompose the reconstructed 
spectrum to mutually independent band-powers keeping resolution as high as possible. 
 
The purpose of this work is to apply the cosmic inversion method to 
the five-year WMAP temperature anisotropy spectrum \cite{WMAP5}, update the reconstructed 
primordial curvature spectrum, and perform the above-mentioned 
band-power analysis by diagonalizing the covariance matrix. 
Then, we revisit the possibility of fine structures in the primordial spectrum. 
Because of the arbitrariness of the primordial spectrum, we inevitably incorporate 
infinite degree of freedom to our analysis, which results in degeneracy 
among spectral shape and cosmological parameters \cite{KNN01,SBKET}. 
In this paper, 
we consider the concordance adiabatic $\Lambda$CDM model, 
where the cosmological parameters (except for the ones characterizing the primordial spectrum) 
are those of the WMAP team's best-fit power-law model \cite{WMAP5ONLY,WMAP5COSMO}, 
and instead focus on the detailed shape of the primordial spectrum. 
Note that, as shown in \cite{KMSY04}, different choices of 
cosmological parameters affect only the overall shape. 
The fine structures of the reconstructed power spectrum 
remain intact in the relatively small wavenumber range we probe. 

This paper is organized as follows: 
In Sec. II, the overview of our analysis is described. 
In Sec. III, we show the reconstructed primordial power spectrum 
from the five-year WMAP data and discuss its implication. 
Finally, Sec. IV is devoted to the conclusion. 

\section{Inversion method}

\subsection{Basic formulas}

Before presenting the inversion method we first list basic formulas 
to be inverted.  Although we only express formulas related to temperature
anisotropy here, the same procedure can be repeated to polarization
anisotropy as well \cite{KSY04}, 
which will give us additional information in the future. 

The temperature anisotropy of photons coming from direction
$\vecnt$ observed at $\vecx$ is decomposed to Fourier modes
and multipole moments as
\begin{eqnarray}
  \frac{\delta T}{T}(\eta,\vecx,\vecnt)
&=& \int\frac{d^3k}{(2\pi)^3} \Theta_{\vecsk}(\eta,\mu)e^{i\vecsk\cdot\vecsx} \nonumber\\
&=& \sum_{\ell,m}a_{\ell m}(\eta,\vecx)Y_{\ell m}(\vecnt),
\end{eqnarray}
where $\mu=\veck\cdot\vecnt \hspace{0.1cm} /|\veck| \equiv \hat{\veck}\cdot\vecnt$ 
and $\eta$ is the conformal time.
In terms of multipole moment in the Fourier space, 
$\Theta_{\ell \vecsk}(\eta)$, which is
defined by
\beq
  \Theta_{\vecsk}(\eta,\mu)=\sum_{\ell} (-i)^\ell 
\Theta_{\ell \vecsk}(\eta)
 P_\ell(\mu),
\eeq
$a_{\ell m}$ is expressed as
\beq
 a_{\ell m}(\eta,\vecx)=\int \frac{d^3k}{(2\pi)^3}
 \Theta_{\vecsk}(\eta,\mu)\frac{4\pi}{2\ell +1}
Y_{\ell m}^\ast(\hat{\veck})
 e^{i\vecsk\cdot\vecsx}.
\eeq
Thanks to the assumption that there exist only adiabatic fluctuations,
we can define the transfer function,  $X_\ell(k)$, from the Fourier mode
of primordial comoving curvature perturbation, 
$\calrk$, to 
$\Theta_{\ell \vecsk}(\eta_0)$
by $\Theta_{\ell \vecsk}=X_\ell(k)\calrk$, where $\eta_0$ is the
present conformal time.
Then the angular power spectrum,
 $C_\ell\equiv\langle |a_{\ell m}|^2\rangle$,
of temperature anisotropy and the power spectrum of curvature
perturbation, $P(k)\equiv \langle |\calrk|^2\rangle$,
are related by
\begin{eqnarray}
\frac{2\ell +1}{4\pi}C_\ell 
&=& \int\frac{d^3k}{(2\pi)^3} \frac{\langle |\Theta_{\ell\vecsk}(\eta_0)|^2\rangle}{2\ell +1} \nonumber\\
&=& \int\frac{d^3k}{(2\pi)^3}  \frac{|X_\ell(k)|^2}{2\ell +1}P(k). 
\label{CL}
\end{eqnarray}
This is the master equation we wish to invert.  Note that $X_\ell(k)$
depends on the cosmological parameters, too.

\subsection{Cosmic inversion method}
Let us introduce the cosmic inversion formula 
which relates the observational CMB anisotropy spectrum 
to the primordial curvature fluctuation spectrum 
by a first-order differential equation. 
Working in the longitudinal gauge,
\beq
  ds^2=a^2(\eta)\lkk -(1+2\Psi(x))d\eta^2+(1+2\Phi(x))d\vex^2\rkk,
\eeq
the Boltzmann equation for $\Theta_\vecsk(\eta,\mu)$ 
can be transformed into the following integral form  \citep{HS95}. 
\begin{eqnarray}
\Theta_\vecsk(\eta_0,\mu)+\Psi_\vecsk(\eta_0) 
= \int_0^{\eta_0} \! d\eta \, \Bigl\{ \lkk\Theta_{0\vecsk}+\Psi_\vecsk 
-i\mu V_{{\rm b}\vecsk}\rkk {\cal V}(\eta) \nonumber\\
 + \lkk\dot{\Psi}_\vecsk-\dot{\Phi}_\vecsk\rkk e^{-\tau(\eta)} \Bigl\} e^{ik\mu(\eta-\eta_0)}, \hspace{1.0cm}
\label{THETA}
\end{eqnarray}
where the overdot denotes the derivative with respect to the 
conformal time. 
Here, $\Psi_\vecsk$ and $\Phi_\vecsk$ are the gauge-invariant quantities
representing the Fourier transform of the
Newton potential 
and the spatial curvature perturbation in this gauge, 
respectively  \citep{B80,KS84}, and 
\begin{eqnarray}
{\cal V}(\eta) \equiv \dot{\tau} e^{-\tau(\eta)}, \quad
    \tau(\eta) \equiv \int_{\eta}^{\eta_0} \! \dot{\tau} d\eta,
\label{VIS&TAU}
\end{eqnarray}
are 
the visibility function and the optical depth for Thomson scattering, 
respectively. 
In the limit that 
the thickness of the last scattering surface (LSS) is negligible, 
we find
${\cal V}(\eta) \approx \delta(\eta-\eta_*)$
and $e^{-\tau(\eta)} \approx \theta(\eta-\eta_*)$
where $\eta_*$ is the recombination time 
when the visibility function is maximal  \cite{HS95}. 
Taking the thickness of the LSS into account, 
we have a better approximation for Eq.~(\ref{THETA}) as 
\begin{eqnarray}
\Theta_\vecsk(\eta_0,\mu)+\Psi_\vecsk(\eta_0)
\approx
\int_{\eta_{*{\rm start}}}^{\eta_{*{\rm end}}} \! d\eta \, \Bigl\{
 \lkk\Theta_{0\vecsk}+\Psi_\vecsk-i\mu V_{{\rm b}\vecsk} \rkk {\cal V}(\eta) \nonumber\\
 +\lkk\dot{\Psi}_\vecsk-\dot{\Phi}_\vecsk\rkk e^{-\tau(\eta)}
 \Bigl\} e^{ik\mu d}, \hspace{1.0cm} \nonumber\\
\equiv  \Theta^{{\rm app}}_\vecsk(\eta_0,\mu)+\Psi_\vecsk(\eta_0), \hspace{2.15cm}
\label{APPROX}
\end{eqnarray}
where $d \equiv \eta_0-\eta_*$ is the conformal distance 
from the present to the LSS and 
$\eta_{*{\rm start}}$ and $\eta_{*{\rm end}}$ are the time 
when the recombination starts and ends, respectively. 
Here, we introduce the transfer functions, $f(k)$ and $g(k)$, defined by 
\begin{eqnarray}
f(k) \calrk \equiv \int_{\eta_{*{\rm start}}}^{\eta_{*{\rm end}}} \! d\eta \, \Bigl\{
 \lkk\Theta_{0\vecsk}(\eta)+\Psi_\vecsk(\eta)\rkk{\cal V}(\eta) \nonumber\\
 +\lkk\dot{\Psi}_\vecsk(\eta)-\dot{\Phi}_\vecsk(\eta)\rkk  e^{-\tau(\eta)} \Bigl\} , \\
g(k) \calrk \equiv 
\int_{\eta_{*{\rm start}}}^{\eta_{*{\rm end}}} \! d\eta \,
\Theta_{1\vecsk}(\eta){\cal V}(\eta) \,. \hspace{1.67cm}
\label{TRANS}
\end{eqnarray}
We can calculate $f(k)$ and $g(k)$ numerically, 
which depend only on the cosmological parameters, 
for we are assuming that only adiabatic fluctuations are present. 

Then, we find the approximated multipole moments as 
\begin{eqnarray}
\frac{\Theta_{\ell\vecsk}^{{\rm app}}(\eta_0)}{2\ell+1}
= \left[ f(k) j_\ell(kd)+g(k) j'_\ell(kd) \right] 
\calrk,
\label{TAPP}
\end{eqnarray}
and the approximated angular correlation function as 
\begin{eqnarray}
C^{{\rm app}}(r)
=\sum_{\ell=\ell_{{\rm min}}}^{\ell_{{\rm max}}} \frac{2\ell+1}{4\pi}
 C^{{\rm app}}_\ell P_\ell \left( 1-\frac{r^2}{2d^2} \right),
\label{CRAPP}
\end{eqnarray}
where $C^{{\rm app}}_\ell$ is obtained 
by putting Eq.~(\ref{TAPP}) into Eq.~(\ref{CL}), 
$r$ is defined as $r=2d \sin(\theta/2)$ on the LSS, 
and $\ell_{{\rm min}}$ and $\ell_{{\rm max}}$ 
are lower and upper bounds on $\ell$ 
due to the limitation of the observation. 
In the small scale limit $r \ll d$, 
using the Fourier sine formula, 
we obtain a first-order differential equation 
for the primordial spectrum of the curvature perturbation, 
\begin{eqnarray}
-k^2f^2(k)P'(k)+\left[ -2k^2f(k)f'(k)+kg^2(k) \right] P(k) \nonumber\\
=4\pi\! \int_0^\infty \!\!\!\!\! dr \,
 \frac{1}{r} \frac{\partial}{\partial r} \{ r^3 C^{{\rm app}}(r) \} \sin kr
 \equiv \Xi (k). \nonumber \\
\label{DIFF}
\end{eqnarray}
Since $f(k)$ and $g(k)$ are oscillatory functions around zero, 
we can find values of $P(k)$ at the zero-points of $f(k)$ as 
\begin{eqnarray}
P(k_s)=\frac{\Xi(k_s)}{k_s \, g^2(k_s)} 
\quad {\rm for} \quad f(k_s)=0,
\label{BC}
\end{eqnarray}
assuming that $P'(k)$ is finite at the singularities, $k=k_s$. 
If the cosmological parameters and the angular power spectrum are given, 
we can solve Eq.~(\ref{DIFF}) 
as a boundary value problem between singularities. 

However, because Eq.~(\ref{DIFF}) is derived by adopting 
the approximation (\ref{APPROX}), 
$C^{{\rm app}}_\ell$ is different from 
the exact angular spectrum $C^{{\rm ex}}_\ell$ for the same initial spectrum. 
The errors caused by the approximation, 
or the relative differences between 
$C^{{\rm app}}_\ell$ and $C^{{\rm ex}}_\ell$ are as large as about 30\%. 
Thus, we should not use the observed power spectrum 
$C^{{\rm obs}}_\ell$ directly in Eq.~(\ref{CRAPP}). 
Instead, we must find $C^{{\rm app}}_\ell$ that 
would be obtained for the real $P(k)$. 
Although this is impossible in the rigorous sense,
we have found an empirical remedy to find $C^{{\rm app}}_\ell$
corresponding to $C^{{\rm obs}}_\ell$ in the following way.
The crucial observation is that the ratio, 
\begin{eqnarray}
b_\ell \equiv
\frac{C^{{\rm ex}}_\ell}{C^{{\rm app}}_\ell},
\label{RATIO}
\end{eqnarray}
is found to be almost independent of $P(k)$  \citep{MSY03}. 
Using this fact, we first calculate the ratio, 
$b^{(0)}_\ell=C^{{\rm ex}(0)}_\ell/C^{{\rm app}(0)}_\ell$, 
for a known fiducial initial spectrum $P^{(0)}(k)$ 
such as the WMAP team's best-fit power-law spectrum. 
Then, inserting $C^{{\rm obs}}_\ell/b^{(0)}_\ell$, 
which is much closer to the actual $C^{{\rm app}}_\ell$, 
into the source term of Eq.~(\ref{DIFF}), 
we may solve for $P(k)$ with good accuracy. 
We may continue this procedure iteratively.

\subsection{Numerical calculation}
Given an initial condition and cosmological parameters, 
we can calculate the transfer functions $f(k)$ and $g(k)$ numerically. 
Then, with the angular correlation function (or equivalently anisotropy 
spectrum), Eq.(\ref{DIFF}) is solved as a boundary value 
problem between the neighboring singularities, 
and hence $P(k)$ is reconstructed. Hereafter, 
we treat $A(k) \equiv k^3P(k)$ instead of $P(k)$ itself for the comprehensive display purpose and 
consistency with the common normalization of fluctuation amplitude. 

We adopt the adiabatic initial condition and fiducial cosmological 
parameter set which is the WMAP team's best-fit power-law model \cite{WMAP5ONLY,WMAP5COSMO} 
to calculate the transfer functions. 
For the reconstruction from the five-year WMAP data, 
the cosmological parameters are $h=0.724$, 
$\Omega_b=0.0432$, $\Omega_\Lambda=0.751$, $\Omega_m=0.249$, and $\tau=0.089$. 
In this case, the positions of the singularities given by Eq.~(\ref{BC}) 
are $kd \simeq 70, 430, 690, ...$, where $d \simeq 1.42\times10^4{\rm Mpc}$. 
Around the singularities, the reconstructed spectrum has 
large numerical errors that are amplified by the observational errors. 
Using various model power spectra to investigate the accuracy of our inversion formula, 
we found that we can achieve the reconstruction 
with good accuracy in the limited range $120 \lesssim kd \lesssim 380$ or 
$8.5\times10^{-3}{\rm Mpc}^{-1} \lesssim k \lesssim 2.7\times10^{-2}{\rm Mpc}^{-1}$. 
In this region, the errors due to our inversion method turn 
out to be much smaller than those due to observational errors including the cosmic variance. 

In practice, we cannot take the upper bound of the integration 
in the right-hand side of Eq.~(\ref{DIFF}) to be infinite. 
The integrand in Eq.~(\ref{DIFF}) is oscillating with 
its amplitude increasing with $r$. 
To evaluate the right-hand side of Eq.(\ref{DIFF}) as finite, 
we convolve an exponentially decreasing 
function with a cutoff scale $r_{cut}$ into the integration of 
the Fourier sine transform. 
As the cutoff scale is made larger, 
the rapid oscillations of the integrand with increasing amplitude 
become numerically uncontrollable. 
On the other hand, if the cutoff scale is made smaller, 
the resolution in the $k$-space becomes worse as 
$\Delta k \sim \pi/r_{{\rm cut}}$. 
For both numerical stability 
and resolution in $k$-space, we adopt the optimized cutoff scale 
of $r_{cut} \simeq 0.8d$. 

For implementing the inversion scheme, 
we employ the routines of CMBFAST code \cite{SZ96} to calculate the
transfer functions, but we have modified it to adopt much finer
resolution than the original one in both $k$ and $\ell$ so that 
we can compute the fine structure of angular power spectra accurately.

\subsection{Monte-Carlo simulation}\label{wind}

In order to incorporate observational errors and the cosmic variance
and to obtain mutually independent band-powers, we employ 
Monte-Carlo method to calculate the covariance matrix of the reconstructed power spectrum. 
Producing 50000 realizations of a temperature anisotropy spectrum based on the WMAP team's best-fit power-law model 
whose statistics obey the likelihood function provided by WMAP team with good precision, 
we obtain 50000 realizations of a reconstructed primordial spectrum. 
(The prescription of simulating anisotropy spectra is described in Appendix.) 
About 1000 samples are sufficient for convergence of the covariance matrix introduced below, 
which assures the robustness of our conclusion. 

The covariance matrix of the reconstructed power spectrum is defined by 
\bea
 K_{ij}&\equiv& \frac{1}{\CN}\sum_{\alpha=1}^{\CN} A_\alpha(k_i)
 A_\alpha(k_j)- \frac{1}{\CN}\sum_{\alpha=1}^{\CN} A_\alpha(k_i)
\frac{1}{\CN}\sum_{\beta=1}^{\CN} A_\beta(k_j) \nonumber \\
&\equiv& \la\!\la A_\alpha(k_i)A_\alpha(k_j)\ra\!\ra_{\alpha}
-\la\!\la A_\alpha(k_i)\ra\!\ra_{\alpha}\la\!\la A_\beta(k_j)\ra\!
\ra_{\beta},
\eea
where $A_\alpha(k_i)$ represents the value of the reconstructed power
spectrum at $k=k_i$ in the $\alpha$-th realization, and $\CN=50000$
as mentioned above.  The resultant  $K_{ij}$ is a square matrix with
its dimension equal to the number of sampling points $k_i$, $N$.
When we solve the cosmic inversion equation, (\ref{DIFF}), we 
discretize the relevant range of $k$ to  more than
2400 points.  We calculate
$K_{ii}$ at each point to estimate the error of $A(k)$ there.
In practice, however, the neighboring $k$-modes are strongly correlated
with each other, as mentioned above, and so the number of 
independent modes are much smaller.  Hence we do not need to, and
in fact, should not take so many points to calculate $K_{ij}$.

Since $K$ is a real symmetric matrix, it can be diagonalized by a
real unitary matrix $U$, to yield
\beq
  UKU^\dag = {\rm diag}\lmk \lambda_1, \lambda_2,...,
  \lambda_N\rmk\equiv \Lambda,
\eeq
where $\lambda_i$'s are the eigenvalues of $K$.  We find they are
positive definite as they should be, provided that
 we take $N$ small enough so that neighboring
modes are not degenerate with each other.  In the present case,
we find that if we take $N\lesssim 50$, the covariance matrix
is well behaved in the sense that the following procedure is
possible with positive definite $\lambda_i$ and the well-behaved
window matrix  defined below. 
In terms of 
\beq
\Lambda^{-1/2}\equiv
{\rm diag}\lmk \lambda_1^{-1/2}, \lambda_2^{-1/2},...,
  \lambda_N^{-1/2}\rmk ,
\eeq
 we define inverse square root of $K$
as $K^{-1/2}\equiv U^\dag \Lambda^{-1/2}U$.  

We also define a
window matrix $W$ by
\beq
  W_{ij}=\frac{(K^{-1/2})_{ij}}{\sum_{m=1}^{N}(K^{-1/2})_{im}},
\eeq
which satisfies the normalization condition $\sum_{j=1}^{N}W_{ij}=1$.
Convolving $P_\alpha(k_i)$ with this window function, we define a new
statistical variable $S_\alpha(k_i)$ as
\beq
  S_\alpha(k_i)\equiv \sum_{j=1}^N W_{ij}A_\alpha(k_j),
\eeq
whose correlation matrix is diagonal and reads
\bea
  \la\!\la S_\alpha(k_i)S_\alpha(k_j)\ra\!\ra_{\alpha}
-\la\!\la S_\alpha(k_i)\ra\!\ra_{\alpha}\la\!\la 
S_\beta(k_j)\ra\!\ra_{\beta} \nonumber \\
= (WK^tW)_{ij}
 =\lkk \sum_{m=1}^N (K^{-1/2})_{im}\rkk^{-2}\delta_{ij},
\eea
where $^tW$ denotes a transposed matrix.

Note that in the previous 
band-power analysis of the power spectrum in the literature, 
decomposition into band-powers or wavelets is done by hand without
calculating the covariance matrix; hence they result in 
either undersampling to lack traceability of fine structures
or oversampling which generates unwanted correlations between different modes. 

\section{Results of reconstruction}

\subsection{Reconstructed primordial spectrum}
\begin{figure}
\includegraphics[scale=1.13]{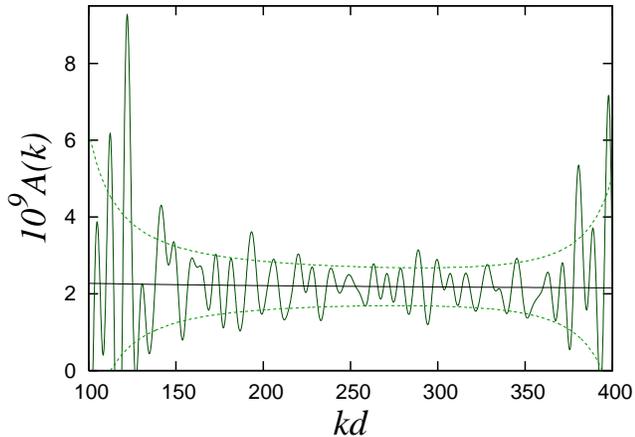}
\caption{
The reconstruction of the primordial spectrum by the cosmic inversion method from 
the five-year WMAP data.  The solid wavy curve is the result of reconstruction
 and the straight line is the best-fit power-law spectrum.  Dotted
 lines are the standard deviation around the best-fit power-law.
}
\label{fig:3rd}
\end{figure}

Figure \ref{fig:3rd} shows the primordial spectrum $A(k)$ reconstructed 
from the five-year WMAP data by the cosmic inversion method. 
In this figure, the solid wavy curve depicts the result of reconstruction 
and the solid straight line is the best-fit power-law 
$\Lambda$CDM model obtained by the five-year WMAP observations, namely, 
the power-law spectrum with $A=2.39\times 10^{-9}$ and $n_s=0.961$, 
where $A$ is the amplitude of curvature perturbation at $k_0=0.002{\rm Mpc}^{-1}$. 
The dotted curves around the power-law are
associated $1\sigma$ standard errors which correspond to the diagonal elements of 
the error covariance matrix (Fig.\ \ref{fig:cov}) calculated by Monte Carlo simulation. 

The modulations of the reconstructed spectrum roughly fit inside the $1\sigma$ borders; 
therefore this oscillatory deviation from the power-law spectrum 
is attributed to the fact that we are analyzing only one random realization of an ensemble
of a simple power-law spectrum. It is not necessarily required that the inflation model
responsible for creation of our Universe should predict such a highly
nontrivial spectrum, for our Universe is merely one of the realizations
of quantum ensemble that is accompanied by significant fluctuations.

Once the possibility of prominent dips around $kd\simeq 180,~220$ and 350 
was pointed out in the analysis of first-year WMAP data \cite{KMSY04}. 
As seen in Fig.\ \ref{fig:3rd}, the modulations on such scales
are fairly degraded in the reconstructed spectrum from the five-year data. 
Since many of the glitches and bites seen in the first-year WMAP anisotropy spectrum 
have disappeared in the five-year WMAP spectrum, the updated primordial spectrum is much more smoothed. 

\subsection{Restored anisotropy spectrum}
\begin{figure}
\includegraphics[scale=1.15]{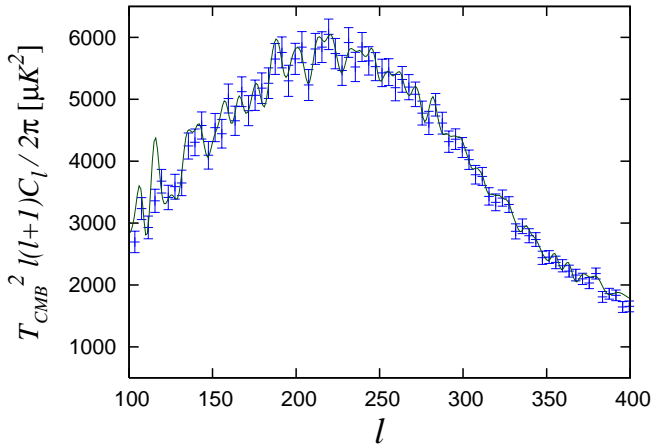}
\caption{The comparison of the binned five-year WMAP data of $\Delta \ell = 4$ 
with the anisotropy spectrum restored from the reconstructed primordial spectrum. 
}
\label{fig:clbin}
\end{figure}

Figure \ref{fig:clbin} illustrates the CMB temperature anisotropy spectrum 
which we restored from the updated primordial spectrum 
by adopting the cosmological parameters of the best-fit power-law model and 
grafting the best-fit power-law spectrum outside of the investigated range. 
The effective $\chi^2$ value in the range of $120 \le \ell \le 380$ for this restored anisotropy spectrum, 
which we calculate using the likelihood tool 
provided by the WMAP team \cite{LAMBDA}, is 246, 
while that for the best-fit power-law model is 273. 
Although the degree of fit is improved significantly,
it does not have the original statistical 
meaning because we have incorporated a functional degree of freedom. 

\subsection{Band-power analysis}

\begin{figure}
\includegraphics[height=8cm,width=8cm]{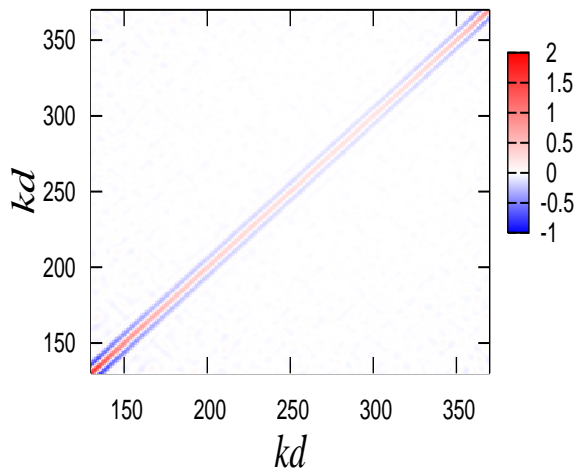}
\caption{
The error covariance matrix of the reconstructed primordial spectrum, $10^9A(k)$, 
calculated by Monte Carlo simulation based on the cosmic inversion method. 
The central strip is indicating positive correlation 
and the surrounding strip is indicating negative correlation. 
Error enhancement on the largest scales of our analysis comes from 
the numerical instability due to the singularity of transfer function 
while error regain on the small scale end is due to 
the domination of measurement error. 
}
\label{fig:cov}
\end{figure}

Despite the fact that our reconstructed power spectrum restores the
fine structures of the observed angular power spectrum of CMB anisotropy well, 
the oscillations observed in the raw reconstructed spectrum may not have true statistical meaning. 
As mentioned in the previous section and we can see in the covariance matrix Fig.\ \ref{fig:cov} explicitly, 
neighboring $k$-modes of the reconstructed power spectrum 
are correlated with each other up to the range $\Delta kd \sim 10$. 
The origin of the correlation is 
the convolution with the transfer function $X_\ell(k)$ between $A(k)$ and $C_\ell$. 
In particular, mapping the primordial spectrum to $C_{\ell}$ erases the information 
that is responsible for the fine structure 
whose characteristic scale is much below the correlation width of $\Delta kd \sim 10$. 

\begin{figure}
\includegraphics[scale=1.15]{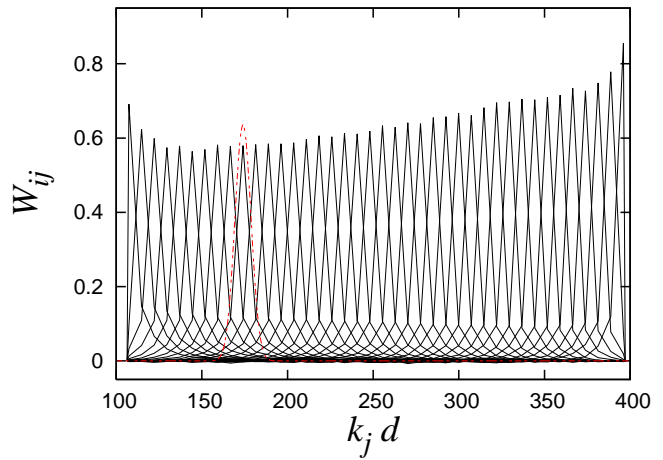}
\caption{The decorrelated window functions $W_{ij}$ for the band-powers. 
The dotted line is a Gaussian fitting of the $i=10$ mode. 
}
\label{fig:window}
\end{figure}

It is important to decompose the power spectrum
into mutually independent modes for evaluating the statistical significance of 
the oscillations in the reconstructed spectrum. 
Here we construct band-powers using the window matrix $W$ defined in 
Sec. \ref{wind}, which diagonalizes the covariance matrix and gives
mutually independent errors. We take the dimension of the window matrix
to $N=40$ and decompose $A(k)$ in the region $100\leq kd \leq 400$ to 40 band-powers.  
Figure \ref{fig:window} shows the window functions for each mode. 

\begin{figure}
\includegraphics[scale=1.13]{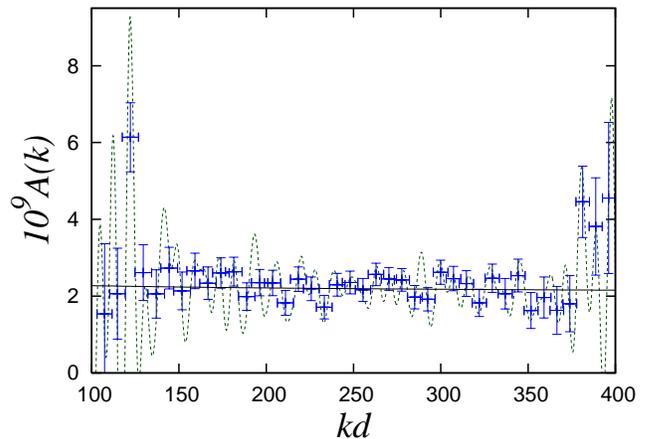}
\caption{
The band-power decomposition of the primordial spectrum
obtained by the cosmic inversion method from the five-year WMAP data.
The horizontal bar indicates the effective width of each window function 
which depicts the dispersion of the fitted Gaussian distribution.
}
\label{fig:A}
\end{figure}

Figure \ref{fig:A} is the result of band-power analysis of the five-year WMAP data. 
In this graph, $i$-th data point indicates the value of 
\beq
 S(k_i) = \sum_{j=1}^N W_{ij} A(k_j),
\eeq
and the vertical error bar represents the variance
\beq
\lkk\la\!\la S_\alpha^2(k_i)\ra\!\ra_{\alpha} -
\la\!\la S_\alpha(k_i)\ra\!\ra_{\alpha}^2\rkk^{1/2} =
\lkk \sum_{m=1}^N (K^{-1/2})_{im}\rkk^{-1}.
\eeq
Here $k_i$ is the location of the peak of the $i$-th line of 
the window matrix $W_{ij}$. 
The horizontal bar, on the other hand, indicates the width of 
the window matrix, where the dispersion of the fitted Gaussian 
is shown (see Fig.\ \ref{fig:window}).  We find their typical 
full width is $\Delta kd \approx 10$ and 
the neighboring horizontal bars barely overlap with each other. 
As is seen clearly in Fig.\ \ref{fig:A}, our band-power reconstruction of the five-year 
power spectrum basically agrees with the best-fit power-law as a whole. 

The $i$-th band-power depends on the multipole moment on the relevant angular scale of $\ell \sim k_id$ and 
also on the surrounding multipoles of $\ell \sim k_id \pm 5$. 
Indeed, we found that the statistics of each band-power subject to Gaussian distribution 
due to the superposition of several multipole moments 
even though the distribution of $C_\ell$ is non-Gaussian (see Appendix). 
By virtue of our band-power analysis, we can also estimate the 
statistical significance of the reconstructed spectrum itself by evaluating the 
deviation from the best-fit power-law spectrum at every band simultaneously. 
We estimated the whole statistical significance of the central 34 bands which correspond to the scales of 
$120 \lesssim kd \lesssim 380$ and found that, 
in terms of reduced $\chi^2$ value which is $0.64$, the reconstructed spectrum from the five-year WMAP data 
is $(-)1.5\sigma$ realization. It is quite consistent with the best-fit power-law spectrum. 

Here the absence of the large modulations around $kd \simeq 180,~220$ and $ 350$ is confirmed again. 
In the band-power analysis, we find dips around
$kd \simeq 180,~220$ and $ 350$ with the statistical significance about 1$\sigma$. 
The nontrivial features reported previously have practically disappeared 
in the reconstructed spectrum from the five-year data. 

Note, however, that in Fig.\ \ref{fig:A}, 
we find a prominent deviation around $kd \simeq 120$ which would be a 
true signal of deviation from a simple power-law spectrum if our
reconstruction method could be trusted there. 
While it is possible that this peak is indeed a real signal associated with the feature observed 
around $\ell\simeq 120$ in the angular power spectrum of the temperature 
anisotropy, it may simply be an artifact because $kd \simeq 120$ is so close 
to the singularity $k_s\simeq 70$ that we can count on our method only marginally there.

\section{Conclusion}

Using the cosmic inversion method, we have reconstructed the primordial
power spectrum of curvature perturbations assuming the absence of
isocurvature modes and the best-fit values of cosmological
parameters for the power-law $\Lambda$CDM model.
While the range of accurate reconstruction is rather narrow,
$120 \lesssim kd \lesssim 380$, we can reproduce the fine structures
of modulations off a simple power-law with which we can recover
the highly oscillatory features observed in $C_\ell$.

The statistical significance of the oscillatory structures in the
reconstructed power spectrum is difficult to quantify due to the
strong correlation among the neighboring $k$-modes.  We have therefore
performed the covariance matrix analysis to calculate the window matrix,
$W$, which diagonalizes the covariance matrix into statistically 
independent modes.  We have chosen the large enough 
number of band-powers,
$N=40$, to probe the feature of the primordial spectrum as precisely as possible while
keeping the overlap of the window functions $W_{ij}$ small enough.

As a result we have found all the independent modes are consistent
with the best-fit power-law $\Lambda$CDM model for the five-year WMAP
data except possibly for a point around $kd\simeq 120$.
Whereas, the possible deviation from a simple power-law around $kd \simeq 180,~220$ and $ 350$ 
reported in the analysis of the first-year WMAP data \cite{KMSY04} has disappeared. 
This difference is due to the fact that the observed $C_\ell$
in the five-year WMAP data has become much smoother than the first-year counterpart around the first peak. 

On the other hand, there still remain some nontrivial deviations from the 
expected  $C_\ell$ of a simple power-law spectrum around $\ell\approx 
 kd \approx 40$ even in the five-year data.  Unfortunately, the cosmic 
inversion method cannot probe the primordial spectrum in this region. 
So we need to develop a different method which can probe the power 
spectrum for smaller wavenumber region $kd \leq 120$ precisely. 

From the covariance matrix analysis, we have shown that statistically 
independent bands of the primordial spectrum have an effective width 
$\Delta kd \approx 10$.  This means that it is difficult to probe the possible 
fine structures on $A(k)$ predicted by, say, trans-Planckian 
processes \cite{Martin:2000xs}, if the scale width of their characteristic 
modulation is narrower than $\Delta kd \approx 10$. 
In other words, even if our result is consistent with a simple power-law 
spectrum, this does not rule out such possibility of narrow modulation of $\Delta kd \ll 10$. 

\acknowledgements
We would like to thank Noriyuki Kogo for his help in numerical
computation. 
We are also 
grateful to Fran\a{c}cois Bouchet, David Spergel, and Masahiro
Takada for useful comments.  This work was partially supported by
JSPS Grant-in-Aid 
for Scientific 
Research Nos.~16340076(JY) and 19340054(JY), JSPS-CNRS Bilateral Joint
Project ``The Early Universe: A Precision Laboratory for High Energy
Physics,'' and JSPS Core-to-Core Program ``International Research
 Network for Dark Energy.'' 

\appendix*
\section{Prescription of generating mock anisotropy spectra}

In order to perform a proper error analysis of the reconstructed power
spectrum, we must prepare a number of realizations of $C_\ell$'s
which obey the correct statistical distribution function.
In the ideal situation with full-sky, uncontaminated observation,
each multipole coefficient $a_{\ell m}$ is Gaussian distributed
and mutually independent
if primordial perturbation is random Gaussian.  Consequently
each angular multipole $C_\ell$ is $\chi^2-$distributed with
$2\ell +1$ degrees of freedom, and different multipoles are
uncorrelated.  In practice, however, reliable observation
can be made in only a finite fraction, $f_{sky}$, of the sky and
different $\ell-$modes are somewhat correlated.  In this situation,
the following form of likelihood function has been proposed 
 \cite{loglike}
and used in the statistical analysis of WMAP data  \cite{WMAP5}.
\bea
 -2\ln\CL =\frac{1}{3}\sum_{\ell,\ell'=\ell_{\min}}^{\ell_{\max}}
(\CC_\ell^D-\CC_\ell^{th})
F_{\ell\ell'}(\CC_{\ell'}^D-\CC_{\ell'}^{th}) \nonumber\\
+\frac{2}{3}\sum_{\ell,\ell'=\ell_{\min}}^{\ell_{\max}}
(\CZ_\ell^D-\CZ_\ell^{th})
F'_{\ell\ell'}(\CZ_{\ell'}^D-\CZ_{\ell'}^{th}), \hspace{0.3cm}
\eea
where $\CC_{\ell}\equiv \ell(\ell+1)C_\ell/2\pi$ and 
$\CZ_{\ell}\equiv \ln({\CC_\ell}+\CN_\ell)$ with
$\CN_\ell \equiv \ell(\ell+1)N_\ell/2\pi$. 
Here $F_{\ell\ell'}$ is the Fisher matrix of $\CC_{\ell}$
and $F'_{\ell\ell'}$ is related with $F_{\ell\ell'}$
as 
\beq
F'_{\ell\ell'}\equiv (\CC_{\ell}^{th}+\CN_\ell)F_{\ell\ell'}
(\CC_{\ell'}^{th}+\CN_{\ell'}). 
\eeq
Superscripts $D$ and $th$ denote
data and the theoretical model, respectively.

The value of each component of Fisher matrix $F_{\ell\ell'}$
is given in  \cite{LAMBDA} where we find
diagonal components
are larger than neighboring off-diagonal elements typically 
by a factor of $\sim 10^2 - 10^3$.  When we depict $C_{\ell}$'s,
these diagonal components of the Fisher matrix are used to indicate
error bars associated with the respective multipole.  
However, since different
multipoles are correlated we would obtain an erroneous result
if we created random samples based on these errors only.
We should first find a basis consisting of linear combinations
of $\CC_{\ell}$'s which diagonalize the Fisher matrix to obtain 
statistically independent quantities.

Here we describe the prescription to find an appropriate basis
to diagonalize the Fisher matrix.  First we note that, 
since the Fisher matrix is a real symmetric matrix, it can be diagonalized by a real unitary matrix which 
we denote by $Q$, namely,
\beq
  Q^{\dag}FQ={\rm diag}(r_{\ell_{\min}},r_{\ell_{\min}+1},
\cdots,r_{\ell_{\max}})\equiv R,
\eeq
where $r_i$'s are eigenvalues of $F_{\ell\ell'}$ and they are all
positive.  In terms of 
\beq
  R^{1/2}\equiv {\rm diag}(r_{\ell_{\min}}^{1/2},r_{\ell_{\min}+1}^{1/2},
 \cdots,r_{\ell_{\max}}^{1/2}),
\eeq
we define matrices
\beq
 V\equiv Q R^{1/2}Q^{\dag}=V^{\dag},
\eeq
and
\beq
 V_D\equiv {\rm diag}(v_{{\ell_{\min}}{\ell_{\min}}},
\cdots,v_{\ell_{\max}\ell_{\max}}),
\eeq
that is, $V$ is a ``square root'' of the Fisher matrix with
$V^2=F$, and $V_D$ is
a diagonal matrix with their diagonal components identical to those
of $V$.
We find
\bea
  (V^{-1}V_D)^{\dag}F(V^{-1}V_D) =V_D^2 \hspace{2.0cm} \nonumber\\
= {\rm diag}(v_{{\ell_{\min}}{\ell_{\min}}}^2,
\cdots,v_{\ell_{\max}\ell_{\max}}^2),
\eea
that is, $F$ can also be diagonalized if it is sandwiched by
$(V^{-1}V_D)^{\dag}$ and $V^{-1}V_D$.  Note, however, that
$v_{\ell\ell}^2$'s are not eigenvalues of $F$ because $V^{-1}V_D$ is
not a unitary matrix.  Nevertheless, defining
\begin{widetext}
\beq
V_D^{-1}V\left(
\begin{array}{c}
\CC_{\ell_{\min}}^D-\CC_{\ell_{\min}}^{th} \\
\CC_{\ell_{\min}+1}^D-\CC_{\ell_{\min}+1}^{th} \\
\vdots \\
\CC_{\ell_{\max}}^D-\CC_{\ell_{\max}}^{th} \\
\end{array}
\right)
=
\left(
\begin{array}{c}
\CB_{\ell_{\min}} \\
\CB_{\ell_{\min}+1} \\
\vdots \\
\CB_{\ell_{\max}} \\
\end{array}
\right),
\eeq
\end{widetext}
and
\begin{widetext}
\beq
 V_D^{-1}V\left(
\begin{array}{c}
(\CC_{\ell_{\min}}^{th}+{\CN}_{\ell_{\min}})(\CZ_{\ell_{\min}}^D-\CZ_{\ell_{\min}}^{th}) \\
(\CC_{\ell_{\min}+1}^{th}+{\CN}_{\ell_{\min}+1})(\CZ_{\ell_{\min}+1}^D-\CZ_{\ell_{\min}+1}^{th}) \\
\vdots \\
(\CC_{\ell_{\max}}^{th}+\CN_{\ell_{\max}})(\CZ_{\ell_{\max}}^D-\CZ_{\ell_{\max}}^{th}) \\
\end{array}
\right)
\equiv \left(
\begin{array}{c}
\tilde{\CB}_{\ell_{\min}} \\
\tilde{\CB}_{\ell_{\min}+1} \\
\vdots \\
\tilde{\CB}_{\ell_{\max}} \\
\end{array}
\right),
\eeq
\end{widetext}
we find the likelihood function is diagonalized as follows.
\beq
  -2\ln\CL=\sum_{\ell={\ell_{\min}}}^{\ell_{\max}} 
\lmk\frac{v^2_{\ell\ell}}{3}
\CB_\ell^2
  +\frac{2v^2_{\ell\ell}}{3}\tilde{\CB}_\ell^2\rmk.  \label{like}
\eeq
Now $\CB_{\ell}$ and $\CB_{\ell'}$ are independent of each other.
On the other hand, $\tilde{\CB}_{\ell}$'s are dependent on
$\CB_{\ell}$'s. 
Reflecting the properties of the Fisher matrix and thanks to the normalization by $V_D^{-1}$, 
we find that each diagonal component of $V_D^{-1}V$ is equal to unity. 
Among the off-diagonal components in each line, $(V_D^{-1}V)_{\ell\ell\pm 2}$ have the largest magnitudes and 
their magnitude is no larger than $0.01$. 
The other off-diagonal components are even smaller.
Therefore, with a good approximation, 
we may put $\CB_\ell\cong \CC_\ell^D-\CC_\ell^{th}$
and $\tilde{\CB}_\ell\cong 
(\CC_\ell^{th}+\CN_\ell)(\CZ_\ell^D-\CZ_\ell^{th})$.  In fact, 
we {\it do not}
adopt these approximations but assume a relation
\beq
  \tilde{\CB}_\ell=(\CC_\ell^{th}+\CN_\ell)\ln\lmk 1+
  \frac{\CB_{\ell}}{\CC_\ell^{th}+\CN_\ell}\rmk, \label{impose}
\eeq
which is inferred from the above approximate relations,
so that each multipole is completely isolated in the likelihood
function (\ref{like}).  With this approximation we can generate
random numbers for each $\ell$ separately obeying a probability
 distribution function (PDF)
\begin{eqnarray}
  \calP [\CB_\ell]d\ln\CB_\ell 
\propto \exp \Bigl[ -\frac{v_{\ell\ell}^2}{6}\CB_\ell^2 \hspace{3.9cm} \nonumber\\
 -\frac{v_{\ell\ell}^2}{3}(\CC_\ell^{th}+\CN_\ell)^2\ln^2\lmk 1+
  \frac{\CB_{\ell}}{\CC_\ell^{th}+\CN_\ell}\rmk \Bigr] d\ln\CB_\ell. \hspace{0.7cm}
 \label{bpdf}
\end{eqnarray}
Without the approximation (\ref{impose}), it would be 
computationally formidable to generate many random samples of an angular power spectrum with the 
appropriate statistical distribution. 

We generate random numbers for each $\CB_\ell$ satisfying the
above PDF (\ref{bpdf}) from which we constitute realizations of
$\CC_{\ell}$ according to
\begin{widetext}
\beq
\left(
\begin{array}{c}
 \CC_{\ell_{\min}}\\
 \CC_{\ell_{\min}+1}\\
 \vdots\\
 \CC_{\ell_{\max}}
\end{array}
\right)
= (V_D^{-1}V)^{-1}
\left(
\begin{array}{c}
 \CB_{\ell_{\min}}\\
 \CB_{\ell_{\min}+1}\\
 \vdots\\
 \CB_{\ell_{\max}}
\end{array}
\right)
+
\left(
\begin{array}{c}
\CC_{\ell_{\min}}^{th}\\
 \CC_{\ell_{\min}+1}^{th}\\
 \vdots\\
 \CC_{\ell_{\max}}^{th}
\end{array}
\right), \label{btoc}
\eeq
\end{widetext}
which have the desired correlation properties and PDFs.

We make 50000 random realizations based on the WMAP team's best-fit power-law angular spectrum 
according to the above prescription and perform inversion using 
the cosmic inversion method to calculate the error covariance matrix of the reconstructed primordial spectrum.


\begin{thebibliography}{}
\bibitem{INFRA}
The infrared cutoff was claimed even before WMAP observation 
and possible resolution had been proposed \cite{JF94,COBE,JY99}.
\end{thebibliography}
\end{document}